\documentclass[preprint,showpacs,preprintnumbers,amsmath,amssymb]{revtex4}
\usepackage{graphicx}
\usepackage{dcolumn}
\usepackage{amssymb}
\usepackage{bm}
\usepackage{epsfig}
\usepackage{psfrag}
\usepackage{ulem}

\def\3dots{\:\raisebox{-0.5ex}{$\stackrel{\textstyle.}{:}$}\:}
\def\beq{\begin{equation}}
\def\eeq{\end{equation}}
\def\bea{\begin{eqnarray}}
\def\eea{\end{eqnarray}}

\begin{document}

\title{Temperature dependent Raman and x-ray studies of spin-ice pyrochlore $Dy_2Ti_2O_7$ and non-magnetic pyrochlore $Lu_2Ti_2O_7$}

\author{Surajit Saha,$^{1}$ Surjeet Singh,$^{2}$ B. Dkhil,$^{3}$, S. Dhar,$^{4}$ R. Suryanarayanan,$^{2}$ G. Dhalenne, $^{2}$ A. Revcolevschi,$^{2}$}

\author{A. K. Sood$^{1}$}%
 \email{asood@physics.iisc.ernet.in, Phone: +91-80-22932964}
\affiliation{$^{1}$Department of Physics, Indian Institute of Science, Bangalore - 560012, India.  
}%

\affiliation{$^{2}$Laboratoire de Physico-Chimie de l'Etat Solide, ICMMO, CNRS, UMR8648, Bat 414, Universite Paris-Sud, 91405 Orsay, France.}%
\affiliation{$^{3}$Laboratoire Structures, Proprietes et Modelisation des Solides, Ecole Centrale Paris, CNRS-UMR8580, Grande voie des vignes, 92295 Chatenay-Malabry Cedex, France}
\affiliation{$^{4}$Tata Institute of Fundamental Research, Mumbai - 400 005, India}


\begin{abstract}

We present here temperature-dependent Raman, x-ray diffraction and specific heat studies between room temperature and 12 K on single crystals of spin-ice pyrochlore compound $Dy_2Ti_2O_7$ and its non-magnetic analogue $Lu_2Ti_2O_7$. Raman data show a ``new'' band not predicted by factor group analysis of Raman-active modes for the pyrochlore structure in $Dy_2Ti_2O_7$, appearing below a temperature of $T_c=$110 K with a concomitant contraction of the cubic unit cell volume as determined from the powder x-ray diffraction analysis. Low temperature Raman experiments on O$^{18}$-isotope substituted $Dy_2Ti_2O_7$ confirm the phonon origin of the ``new'' mode. These findings, absent in $Lu_2Ti_2O_7$, suggest that the room temperature cubic lattice of the pyrochlore $Dy_2Ti_2O_7$ undergoes a ``subtle'' structural transformation near $T_c$. We find anomalous \textit{red-shift} of some of the phonon modes in both the $Dy_2Ti_2O_7$ and the $Lu_2Ti_2O_7$ as the temperature decreases, which is attributed to strong phonon-phonon anharmonic interactions.

\end{abstract}


\maketitle

\section{Introduction}

Geometrically frustrated magnetic systems are those where spins cannot simultaneously minimize all their pairwise exchange energies because of geometrical reasons and hence cannot find a unique ground state \cite{ARMS-24-453, PT-59-24}. In a real system, however, any small perturbation to these interactions can favor a particular ground state thus lifting the macroscopic degeneracy. The perturbation can be in the form of single-ion anisotropy, next or higher neighbor interactions, thermal or quantum fluctuations, dipolar interactions, lattice deformations, or externally applied stimuli such as pressure or magnetic field. The pyrochlore structure (space group $Fd\bar{3}m$), with stoichiometry $A_2B_2O_7$, is a typical example of geometrically frustrated magnets in three dimensions. In this structure, $A^{3+}$ and $B^{4+}$ ions form distinct interpenetrating networks of corner-sharing tetrahedra, displaced from each other by half the unit cell dimension. If either A or B ion has a magnetic moment, strong geometrical frustration results, which in the presence of one or more perturbative terms, gives rise to various interesting ground states. Such is the case for the family of insulating rare-earth titanates ($R_2Ti_2O_7$; R = trivalent rare-earth ion), which show a wide diversity in their magnetic behavior (For a recent review on these pyrochlores, see ref \cite{JAC-408-444}) from a N\'{e}el ordered state in $Er_2Ti_2O_7$ \cite{PRB-68-020401} to a spin-liquid state in $Tb_2Ti_2O_7$ \cite{PRL-82-1012} and spin-ice states in $Dy_2Ti_2O_7$ \cite{Nature-399-333} and $Ho_2Ti_2O_7$ \cite{PRL-79-2554}. In the spin-ice ground state, two spins on an elementary tetrahedron point in and the remaining two point out, an arrangement that is analogous to protons ordering in ordinary water ice \cite{book}. Consequently, the ground state degeneracy and hence the ``zero-point'' entropy of a spin-ice is shown \cite{Nature-399-333} to be very close to what Linus Pauling estimated for the water ice nearly 50 years ago \cite{book}. Since spins can easily be manipulated by external stimuli, such as magnetic field \cite{PRB-65-054410, PRL-90-207205}, studying spin ice is of considerable importance even from the point of view of understanding the properties of ordinary water ice. For these reasons, the spin ices $Dy_2Ti_2O_7$ and $Ho_2Ti_2O_7$ have attracted considerable recent attention. Besides these two canonical spin ices, other examples of spin-ice or spin-ice like ground state include $Ho_2Sn_2O_7$ \cite{JPCM-12-L649} and $Tb_2Sn_2O_7$ \cite{PRL-94-246402} members in the analogous stannate series, and, $Nd_2Mo_2O_7$ \cite{Science-291-2573} and $Sm_2Mo_2O_7$ \cite{PRB-77-020406} in the molybdate pyrochlore series.

Spin-ice state in $Dy_2Ti_2O_7$ and $Ho_2Ti_2O_7$ emerges at very low temperature when the dipolar interactions between the large Ising-like $Dy^{3+}(Ho^{3+})$ spins overwhelm the near-neighbor antiferromagnetic exchange, and therefore, a majority of studies on these pyrochlores, until recently, were limited to the liquid Helium or lower temperature range. Recently, vibrational properties of some titanate pyrochlores (including the spin-ice $Dy_2Ti_2O_7$) have been studied using infrared reflectivity \cite{JPCM-17-5225} and Raman spectroscopy \cite{JRS-39-537, PRB-77-214310} up to room temperature. In the work by M\c{a}czka \textit{et al.} \cite{JRS-39-537} on $R_2Ti_2O_7$ (R = Er, Gd and Dy), Raman studies of $Gd_2Ti_2O_7$ and $Er_2Ti_2O_7$ were carried out to determine the temperature dependence of Raman mode frequencies and line-widths. These authors, however, could not follow the temperature dependence of Raman modes in $Dy_2Ti_2O_7$. These studies have shown that mode frequencies of a few phonons decrease upon cooling, a behavior not often observed in solids and hence termed as ``anomalous''. This anomalous behavior has been qualitatively attributed by Bi \textit{et al.} \cite{JPCM-17-5225} to strong spin-phonon coupling in $Dy_2Ti_2O_7$ for a few modes. Since the strength of exchange interaction is relatively much weaker for titanate pyrochlores (ranging between $\sim$ 19 K in $Tb_2Ti_2O_7$ \cite{PRL-83-211, PRB-64-224416} to $\sim$ 0.25 K in $Sm_2Ti_2O_7$ \cite{PRB-77-054408}), and there is no long-range spin order in $Dy_2Ti_2O_7$ as well as in $Gd_2Ti_2O_7$ and $Er_2Ti_2O_7$ down to very low temperatures \cite{JAC-408-444}, this qualitative explanation of the anomalous temperature dependence in terms of spin-phonon coupling has been questioned \cite{JRS-39-537}. Another explanation of this anomalous behavior in terms of a coupling between phonons and crystal field excitations can also be ruled out because $Gd_2Ti_2O_7$ also exhibit comparable phonon anomalies even though in $Gd^{3+}$ ion no crystal field effect is expected due to spherical symmetry of the 4f charge cloud. It is suggested in ref. \cite{JRS-39-537} that the anomalous temperature dependence could arise from anharmonic interactions. In order to understand the anomalous temperature dependence of phonon modes, we have done detailed Raman studies on single crystals of $Dy_2Ti_2O_7$ and its non-magnetic analogue $Lu_2Ti_2O_7$ over a temperature range of 12 to 300 K. A quantitative comparison of the phonons of these two pyrochlores is done to understand the role of spin-phonon coupling in giving rise to the anomalous phonon behaviors. Most interestingly, we show here that an equally anomalous phonon behavior is present for the non-magnetic $Lu_2Ti_2O_7$ as well. Another important outcome of our study is the observation of a ``new'' mode near 300 cm$^{-1}$ in the Raman spectra of $Dy_2Ti_2O_7$ recorded below T  = 110 K. Recently, Lummen \textit{et al.} \cite{PRB-77-214310} also made a similar observation, who tentatively assigned the ``new'' mode to the first excited crystal field level of Dy$^{3+}$. By performing Raman spectrscopic experiments on an O$^{18}$-isotope enriched crystal of $Dy_2Ti_2O_7$, we show here  for the first time that the ``new'' mode is actually a phonon mode not expected from the factor group analysis of Raman active modes for the space group $Fd\bar{3}m$ of the pyrochlore structure. We also show that the emergence of ``new'' mode is corroborated by an abrupt change in the temperature dependence of the lattice constant by powder x-ray diffraction studies. These findings, absent in $Lu_2Ti_2O_7$, suggest a possible lattice symmetry lowering in $Dy_2Ti_2O_7$ below 110 K.

\section{Experimental Techniques}

Stoichiometric amounts of $Dy_2O_3/Lu_2O_3$ (99.99 $\%$) and $TiO_2$ (99.99 $\%$) were mixed thoroughly and heated at 1200 $^\circ$C for about 15 h. The resulting mixture was well ground and isostatically pressed into rods of about 8 cm long and 5 mm diameter. These rods were sintered at 1400 $^\circ$C in air for about 72 h. This procedure was repeated until the compound $Dy_2Ti_2O_7/Lu_2Ti_2O_7$ was formed, as revealed by powder x-ray diffraction analysis, with no traces of any secondary phase. These rods were then subjected to single crystal growth by the floating-zone method in an infrared image furnace under flowing oxygen. X-ray diffraction, carried out on the powder obtained by crushing part of a single crystalline sample and energy dispersive x-ray analysis on a scanning electron microscope indicated a pure pyrochlore $Dy_2Ti_2O_7/Lu_2Ti_2O_7$ phase. The L\"{a}ue back-reflection technique is used to orient the crystal along the principal crystallographic directions. To prepare the $O^{18}$-isotope rich samples of $Dy_2Ti_2O_7$, thin crystal slices of $Dy_2Ti_2O_7$ (O$^{16}$) were cut from their as-grown boule. The thickness was further reduced to less than 500 $\mu$m by polishing the flat ends. Each crystal slice thus obtained was cut into two halves. While one half was annealed in oxygen atmosphere enriched with $^{18}$O$_2$ (95 \%) isotope at 950 $^0$C for 7 days, the other half was subjected to an identical heat treatment but under normal oxygen atmosphere. We found that the samples annealed under $^{18}$O$_2$ atmosphere gained nearly 2 \% in weight, while the weight of samples annealed under normal oxygen atmosphere remained unchanged, which indicates that a substantial isotopic exchange took place during the heat treatement.

Raman spectroscopic measurements on a (111) cut thin single-crystalline slice (0.5 mm thick and 3 mm in diameter, polished down to a roughness of almost 10 $\mu$m) of $Dy_2Ti_2O_7$ and $Lu_2Ti_2O_7$ were performed at low temperatures in back-scattering geometry, using 514.5 nm line of an $Ar^+$ ion laser (Coherent Innova 300) with $\sim$ 15 mW of power falling on the sample. Temperature scanning was done using a CTI-Cryogenics Closed Cycle Refrigerator. Temperature was measured and controlled (with a maximum error of 0.5 K) using a calibrated Si-diode sensor and a CRYO-CON 34 temperature controller, respectively. The scattered light was collected by a lens and was analyzed using a computer controlled DILOR XY Raman spectrometer having three holographic gratings (1800 groves/mm) coupled to a liquid nitrogen-cooled charged coupled device, CCD 3000 (Jobin Yvon-SPEX make) of pixel resolution 0.85 cm$^{-1}$. The instrumental broadening is about 5 cm$^{-1}$.

High resolution x-ray diffraction measurements were performed between 10-300 K (with temperature accuracy better than 0.5 K) using a highly accurate two-axis diffractometer in a Bragg-Brentano geometry (focalization circle of 50 $\mu$m) using the Cu-K$_{\beta}$ line ($\lambda$=1.39223 $\AA $) of a 18 kW rotating anode.

Specific heat of $Dy_2Ti_2O_7$ and $Lu_2Ti_2O_7$ was measured between 5 K and 300 K using a Quantum Design Physical Property Measurement System (PPMS). Thin single crystal slices (thickness $\sim$ 500 $\mu$m) weighing about 10 mg each were mirror polished and mounted on the sample holder of PPMS. Addenda specific heat of the bare sample holder and a small amount of APIZONE N grease (to make thermal contact with the sample) was measured each time before loading the single crystals on the sample holder.

\section{Results}

\subsection{Raman spectrum of $Dy_2Ti_2O_7$ and $Lu_2Ti_2O_7$}

\begin{figure}
\begin{center}
\leavevmode
\includegraphics[width=0.9\textwidth]{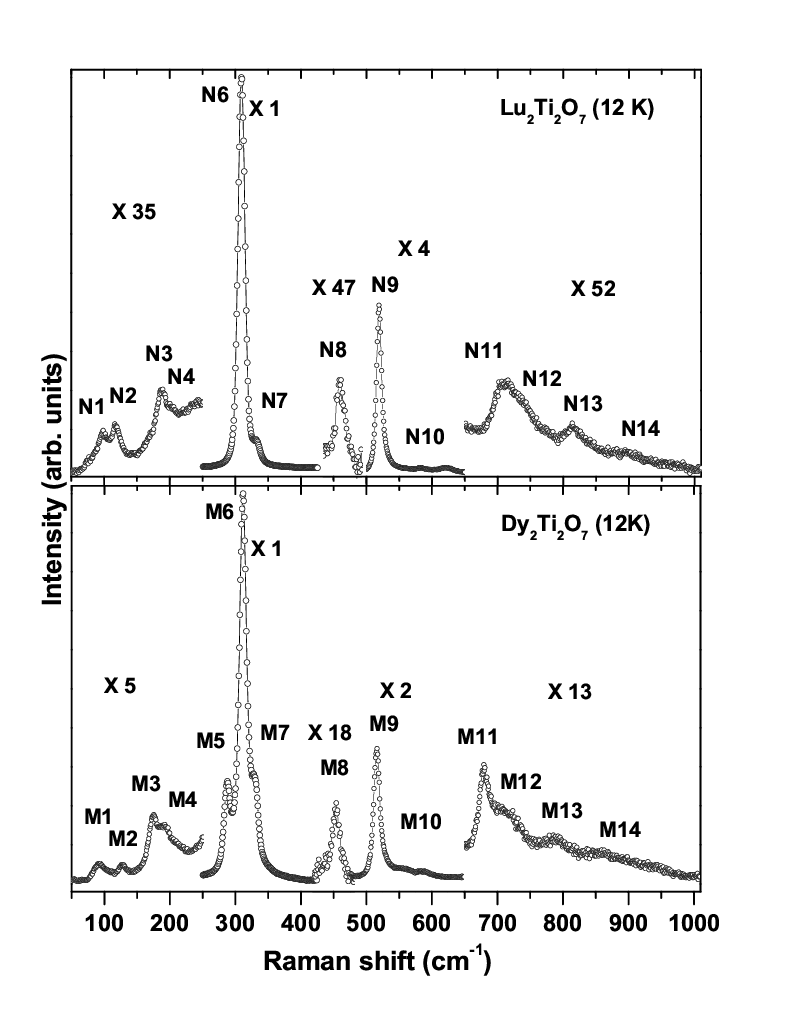} 
\caption{Raman spectrum of $Dy_2Ti_2O_7$ and $Lu_2Ti_2O_7$ at 12 K. Open circles represent the experimental data. Raman intensity in the regions from 50-250 $cm^{-1}$, 425-475 $cm^{-1}$, 475-650 cm$^{-1}$ and 650-1010 $cm^{-1}$ are rescaled with respect to the region between 250-425 $cm^{-1}$ in order to view the weak modes clearly. The different modes labeled as M1(N1) to M14(N14) are assigned in the text (Table-I).} \label{Fig:1}
\end{center}
\end{figure}

The pyrochlore family belongs to the space group $Fd\bar{3}m (O^{h}_{7})$ with an $A_2B_2O_6O^{\prime}$ stoichiometry, where $A^{3+}$ occupies the 16d and $B^{4+}$ occupies the 16c positions and the oxygen atoms O and O$^{\prime}$ occupy the 48$f$ and 8$b$ sites, respectively. Factor group analysis for this family of structures suggests six Raman active modes ($A_{1g}+E_g+4F_{2g}$) and seven infrared active modes ($7F_{1u}$). Fig. \ref{Fig:1} (bottom panel) shows the Raman spectrum of $Dy_2Ti_2O_7$ at 12 K between wavenumber 50 and 1000 cm$^{-1}$ which is fitted with fourteen Lorentzians, labeled M1 to M14. The assignment of various modes have been done following the previously reported results of ref. \cite{CPL-413-248, JPCB-106-4663} (see Table-I for details). The six Raman active modes are labeled M3 (174 cm$^{-1}$, $F_{2g}$), M6 (312 cm$^{-1}$, $F_{2g}$), M7 (330 cm$^{-1}$, $E_{g}$), M8 (453 cm$^{-1}$, $F_{2g}$), M9 (515 cm$^{-1}$, $A_{1g}$) and M10 (563 cm$^{-1}$, $F_{2g}$). The top panel of Fig. \ref{Fig:1} shows the Raman spectrum of $Lu_2Ti_2O_7$ at 12 K where all the modes are seen, except the mode M5. The six Raman active modes of $Lu_2Ti_2O_7$ appear at 188 cm$^{-1}$ (N3, $F_{2g}$), 313 cm$^{-1}$ (N6, $F_{2g}$), 336 cm$^{-1}$ (N7, $E_g$), 458 cm$^{-1}$ (N8, $F_{2g}$), 520 cm$^{-1}$ (N9, $F_{2g}$) and 609 cm$^{-1}$ (N10, $F_{2g}$). Some of the previous studies \cite{PRB-74-064109, JRS-14-63, JRS-32-41} labeled the mode M9(N9) as $A_{1g}+F_{2g}$, and overlooked the weak mode near 450 cm$^{-1}$. We and Lummen \textit{et al.} \cite{PRB-77-214310} have observed a mode near 450 cm$^{-1}$ in several of the pyrochlore titanates. This mode has sometimes been assigned to $TiO_2$ as an impurity phase in the pyrochlore sample \cite{PR-154-522}. We rule out this possibility because a strong mode of $TiO_2$, near 610 cm$^{-1}$, is not seen in our single-crystals. We have therefore, assigned the 450 cm$^{-1}$ mode to $F_{2g}$ phonon, as done in ref. \cite{CPL-413-248, JPCB-106-4663}. 

We now come to the origin of the additional modes M1(N1), M2(N2), M4(N4), M11(N11), M12(N12), M13(N13) and M14(N14), also observed in refs. \cite{JRS-14-63, JRS-39-537} and seen in both magnetic and non-magnetic pyrochlores (Fig. \ref{Fig:1}). The modes labelled as M11 to M14 can be assigned to second order Raman scattering \cite{JRS-39-537}. Since the modes M1, M2 and M4 are seen in both $Dy_2Ti_2O_7$ (magnetic) and $Lu_2Ti_2O_7$ (non-magnetic), these can not be assigned to crystal field transitions of $Dy^{3+}$ in $Dy_2Ti_2O_7$. We, therefore, assign these modes to infra-red active or silent modes rendered Raman active due to a possible lowering of the local symmetry at some crystallographic sites, as suggested in ref. \cite{JRS-14-63}. We have followed the evolution of the different modes with temperature from 12 K to 300 K for both $Dy_2Ti_2O_7$ and $Lu_2Ti_2O_7$ and the results are discussed below.

\subsection{Evolution of Raman spectra with temperature}

\begin{figure}
\begin{center}
\leavevmode
\includegraphics[width=0.6\textwidth]{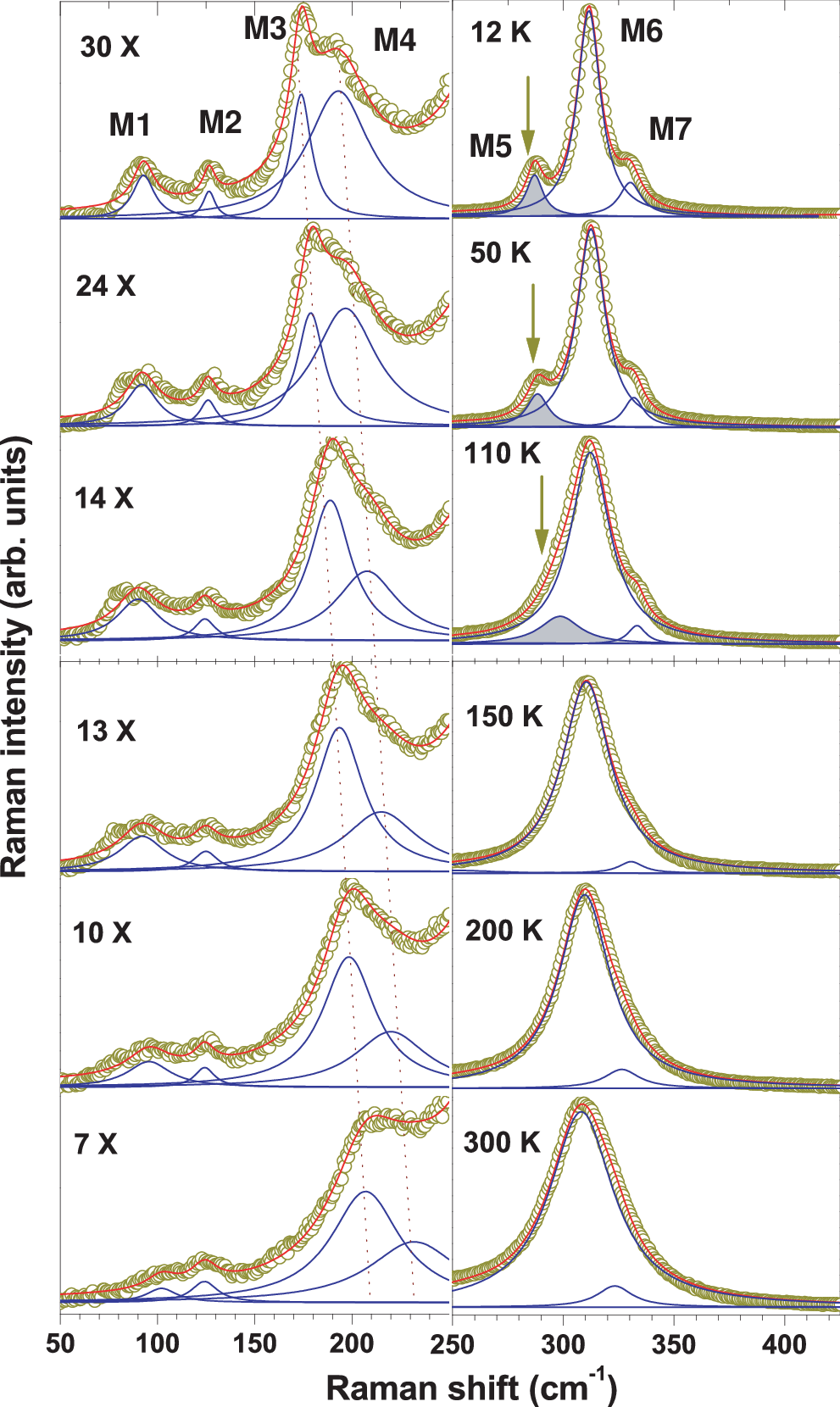}
\caption{(Color online) Raman spectra of $Dy_2Ti_2O_7$ at different temperatures. Solid lines are fit to the experimental data (open circles), as discussed in text. A new mode (M5) appears at 110 K which has been marked by arrows in panels corresponding to 100, 50 and 12 K. Raman intensity in the low frequency regions (50-250 $cm^{-1}$) have been rescaled.} \label{Fig:2}
\end{center}
\end{figure}

\begin{figure}
\begin{center}
\leavevmode
\includegraphics[width=0.8\textwidth]{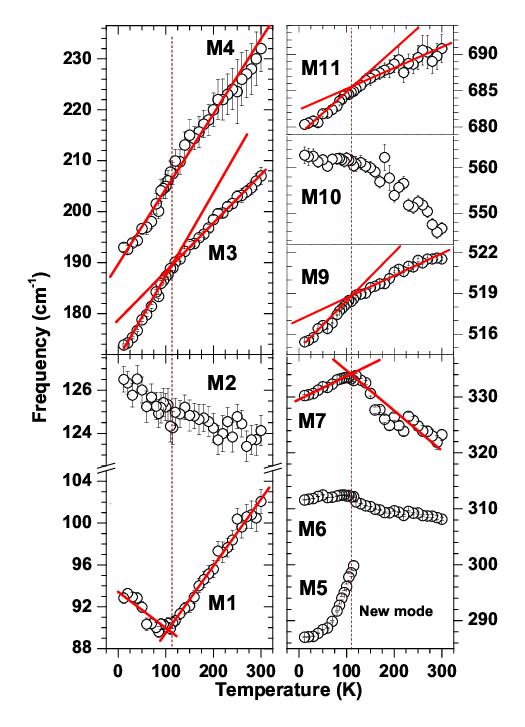}
\caption{(Color online) Temperature dependence of the different modes of $Dy_2Ti_2O_7$. Solid lines are guide to eye. New mode (M5) appears below 110 K} \label{Fig:3}
\end{center}
\end{figure}

Raman spectra of $Dy_2Ti_2O_7$ have been recorded at several temperatures varying from 12 K to 300 K. The maximum changes are seen in Raman bands M1 to M7 in the spectral range of 50 to 400 cm$^{-1}$. We, therefore, show in Fig. \ref{Fig:2} the Raman spectra at a few typical temperatures in this spectral range. The modes M1, M2, M3 and M4 are weak, their intensities are, therefore, rescaled as indicated in Fig. \ref{Fig:2} with respect to the mode M6. It can be clearly seen that for T $>$ 110 K, the spectra in the 250 to 400 cm$^{-1}$ range can be fitted by two Lorentzians representing the two allowed Raman modes (M6 and M7). However, the spectra at T = 110 K and below, indicate a ``new'' mode (termed as M5) marked by an arrow in the panels corresponding to T = 110, 50 and 12 K  in Fig. \ref{Fig:2}. The frequencies of the modes M1, M2, M3, M4, M5 (new mode), M6, M7, M9, M10 and M11 are plotted against temperature in Fig. \ref{Fig:3}. It can be seen that the modes M1, M3, M4, M9, M11 and the new mode M5 (which appears below 110 K) show large anomalous \textit{red-shifts} upon decreasing temperature. The mode M7 initially shows a \textit{blue-shift} and then a small \textit{red-shift} below 110 K. The mode M1 shows a small \textit{blue-shift} as temperature decreases below 110 K. However, the intense mode M6 and the modes M2 and M10 exhibit \textit{blue-shift} with decreasing temperature, an usual behavior seen in solids. The temperature dependence of the M3 and M9 modes show an interesting trend, namely, the rate of softening ($\frac{\partial \omega}{\partial T}$) increases below 110 K.

\begin{figure}
\begin{center}
\leavevmode
\includegraphics[width=0.8\textwidth]{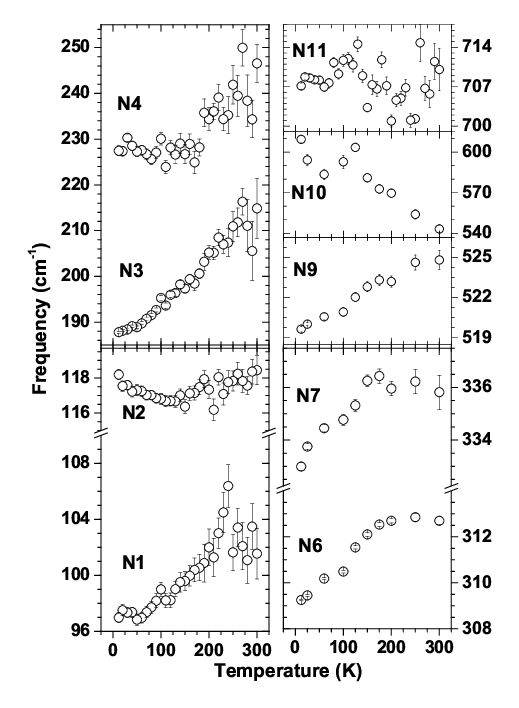}
\caption{Temperature dependence of the different modes of $Lu_2Ti_2O_7$.} \label{Fig:4}
\end{center}
\end{figure}

Temperature dependence of the some of the modes of non-magnetic $Lu_2Ti_2O_7$ is plotted in Fig. \ref{Fig:4}. It is clearly seen that phonon modes N1, N2, N3, N4, N6, N7, and N9 show anomalous temperature dependence, similar to the ones seen in $Dy_2Ti_2O_7$ phonons.

\subsection{Appearance of a new mode in $Dy_2Ti_2O_7$}

In Fig. \ref{Fig:2}, it is clearly evident that a ``new'' mode (M5) emerges near $\omega \sim$ 300 cm$^{-1}$ below $T_c$ = 110 K in $Dy_2Ti_2O_7$ whereas no new Raman mode has been seen in $Lu_2Ti_2O_7$ even at 12 K (Fig. \ref{Fig:1}) in the present study or in other titanate pyrochlores, like, $Gd_2Ti_2O_7$ \cite{JRS-39-537} , $Er_2Ti_2O_7$ \cite{JRS-39-537}, $Ho_2Ti_2O_7$ \cite{PRB-77-214310} and $Sm_2Ti_2O_7$ \cite{PRB-77-054408} studied previously. At room temperature, the Raman spectrum of $Dy_2Ti_2O_7$ between 250 and 400 cm$^{-1}$ is satisfactorily fitted to a sum of two Lorentzian functions centered at 308 and 322 cm$^{-1}$, corresponding to the modes $F_{2g}$ (M6) and $E_g$ (M7), respectively (see Table-I). A similar fit worked equally well at lower temperatures down to about 115 K, as shown for T = 200 and 150 K spectra in the lower panels of Fig. \ref{Fig:2}. However, for the spectrum recorded at $T_c$ = 110 K, it became neccessary to include a third Lorentzian in order to capture an essential feature, which is the emergence of a ``new'' mode on the low-frequency side of the mode M6 (middle panel of Fig. \ref{Fig:2}). This mode gets further resolved and more pronounced as the sample temperature is lowered below $T_c$, as shown by arrows in the panels corresponding to 110, 50 and 12 K. 

\begin{figure}
\begin{center}
\leavevmode
\includegraphics[width=\textwidth]{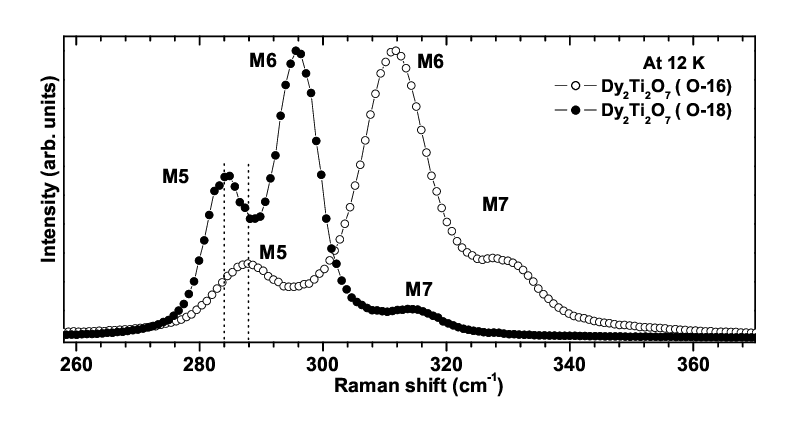}
\caption{Raman spectra of O$^{16}$ (open circle) and O$^{18}$ (closed circle) isotopic $Dy_2Ti_2O_7$ at 12 K in the range of 260 cm$^{-1}$ to 370 cm$^{-1}$. A change in frequencies are seen with a change in O-ion isotopic mass.} \label{Fig:5}
\end{center}
\end{figure}

We note that Lummen \textit{et al.} \cite{PRB-77-214310} and M\c{a}czka  \textit{et al.} \cite{JRS-39-537} also see this ``new'' mode in their low-temperature Raman spectra of $Dy_2Ti_2O_7$. Lummen \textit{et al.} tentatively assigned the ``new'' mode near 300 cm$^{-1}$ to the first excited CF level after the crystal field splitting estimation due to Rosenkranz \textit{et al.} \cite{JAP-87-5914}. However, the crystal field splittings of the lowest $J$-multiplet calculated by Jana \textit{et al.} \cite{JMMM-248-7} differ considerably. Our results of the specific heat measurements on $Dy_2Ti_2O_7$ and $Lu_2Ti_2O_7$, discussed in the later part of this article, suggest that the energy gap between the ground and the first excited CF level is $\sim$ 250 cm$^{-1}$ which is closer to Rosenkranz's estimation than Jana's calculations. Based on these observations, the assignment of the ``new'' mode as a CF mode (transition from ground state to the first excited CF level) by Lummen \textit{et al.} \cite{PRB-77-214310} looks attractive. However, this assignment of the ``new'' mode to be the CF origin is not compatible with our low temperature Raman experiments on single crystal of O$^{18}$-isotopic $Dy_2Ti_2O_7$. Since all the Raman active modes in pyrochlores involve O-ion vibrations only, replacement of O$^{16}$ by O$^{18}$ in $Dy_2Ti_2O_7$ will \textit{red-shift} all the phonon modes, whereas the positions of CF modes, if present, will remain unchanged. Our experimental results shown in Fig. \ref{Fig:5} show a \textit{red-shift} of $\sim$ 4 cm$^{-1}$ of the ``new'' mode when O$^{16}$ is replaced by O$^{18}$, thus suggesting the ``new'' mode to be a lattice vibration and not a CF transition. This assignment is further strengthened by the fact that the full width at half maxima ($\Gamma$) is less in O$^{18}$ rich $Dy_2Ti_2O_7$ ($\Gamma$=6.7 cm$^{-1}$) as compared to its value ($\Gamma$ = 9.1 cm$^{-1}$) in O$^{16}$ rich $Dy_2Ti_2O_7$. This decrease in $\Gamma$ is also seen for the M6 and M7 modes. It has been shown earlier that the phonon linewidth depends on isotopic mass as $\Gamma \propto M_{isotope}^{-1}$ \cite{SSC-117-201}. This, therefore, strongly suggests that the ``new'' mode M5 near 300 cm$^{-1}$ is of phonon origin. Another possibility is that the ``new'' mode (M5) becomes observable below 110 K when the linewidth of the mode M6 decreases. This explanation will not be consistent with our results on $Lu_2Ti_2O_7$: at 300 K the linewidth of the N6 mode in $Lu_2Ti_2O_7$ is similar to the corresponding M6 mode and no ``new'' mode is seen in $Lu_2Ti_2O_7$ at low temperatures. The appearance of the ``new'' mode (M5) is, therefore, attributed to a Raman inactive phonon mode becoming Raman active due to lowering of the site symmetries arising from local structural deformation below 110 K. Following the indication of a possible structural transition, we have performed x-ray diffraction experiments as a function of temperature and the results are discussed below.

\subsection{Low temperature x-ray diffraction}

\begin{figure}
\begin{center}
\leavevmode

\includegraphics[width=0.7\textwidth]{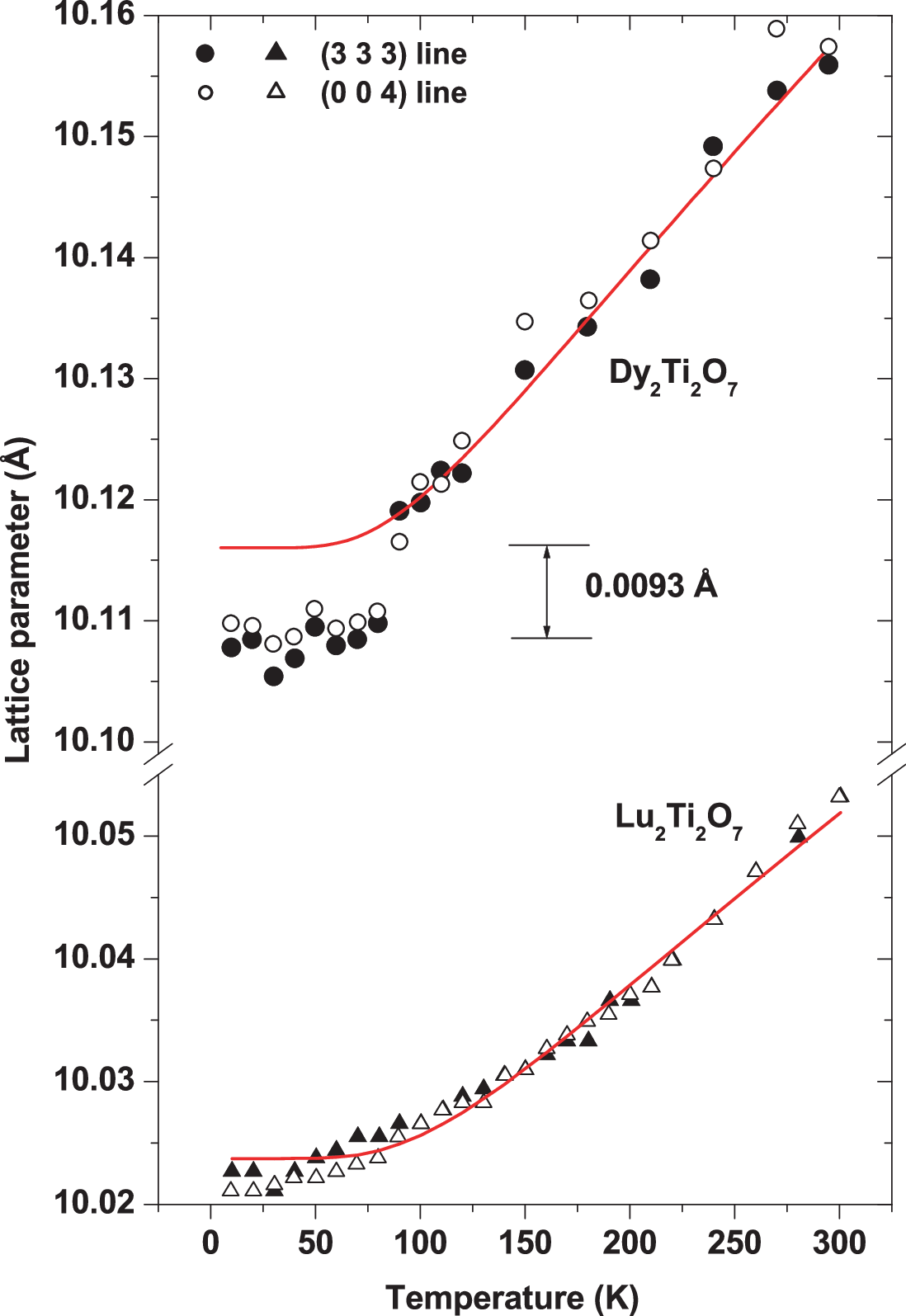}
\caption{(Color online) Variation of the lattice parameter of $Dy_2Ti_2O_7$ with temperature. An abrupt jump of the lattice parameter, indicative of a structural deformation, is seen near 100 K, a temperature where ``new'' mode appears in Raman spectrum. In comparison, the lattice parameter of non-magnetic $Lu_2Ti_2O_7$ varies smoothly over the entire range of temperature. Solid lines are fit to the data as discussed in text.} \label{Fig:6}
\end{center}
\end{figure}

In order to examine if the appearance of the ``new'' mode in $Dy_2Ti_2O_7$ at 110 K is related to a possible structural transition, we have recorded high accuracy powder x-ray diffraction patterns from room temperature down to 10 K. Inspection of the whole diffraction pattern recorded at 10 K did not show additional new peaks nor splitting or broadening of the existing Bragg peaks. However, as shown in Fig. \ref{Fig:6}, the temperature dependence of the lattice parameter ``a'' extracted from (004) and (333) Bragg reflections, clearly displays a discontinuous change in the lattice parameter near 100 K, the temperature at which the new mode (M5) appears in our Raman data. In contrast, the temperature variation of the lattice parameter in its non-magnetic analogue $Lu_2Ti_2O_7$, plotted in Fig. \ref{Fig:6} for comparison, is smooth as expected in the absence of a structural deformation. The solid lines in Fig. \ref{Fig:6} are fit to the data by the relation $a(T)=a_0[1+\frac{be^{c/T}} {T(e^{c/T}-1)^2 }]$, where $a_0$ is the lattice constant at 0 K and b and c are fitting parameters \cite{KittleBook}. For $Dy_2Ti_2O_7$ $a_0$=10.116 \AA, b=3.152 K, c=436.3 K and for $Lu_2Ti_2O_7$ they are 10.024 \AA, 3.119 K and 513.0 K, respectively.

A survey of the literature for experiments on $Dy_2Ti_2O_7$ showing anomalous behavior near 110 K directed us to a report \cite{SSC-12-693} on the Nuclear Gamma Resonance of $^{161}Dy^{3+}$ in $Dy_2Ti_2O_7$ by M\"{o}ssbauer spectroscopy suggesting a change in the nuclear quadrupolar interaction with temperature. The total electric field gradient in the $Dy^{3+}$ nuclei comes from the 4$f$ orbitals of $Dy^{3+}$ and the external charges around, in the lattice. As suggested in ref \cite{SSC-12-693}, there is no significant change in the lattice contribution with temperature but a dramatic change was seen in the electric field gradient due to 4$f$ orbitals with temperature that becomes almost zero at 110 K. We propose a possibility  that when the electric field gradient at $Dy^{3+}$ ions due to their 4$f$ orbitals becomes negligibly small, the surrounding atoms in the lattice slightly readjust and move out of their regular Wyckoff positions thus deforming the local crystal symmetry without disturbing the average electric field gradient due to the surrounding ions in the lattice. There is a recent study on the temperature dependence of spin-relaxation time in $Dy_2Ti_2O_7$ \cite{PRB-75-140402}. Unfortunately, the data are taken only between 10-90 K and at 300 K. The data corresponding to 300 K do not agree with the value extrapolated from the Arrhenius behavior fitted to the data between 10 K and 90 K. Later, experiments by Kitagawa \textit{et al.} \cite{PRB-77-214403} filled up the gap in the data in  the range of 90 K to 300 K. In Fig. 6 of ref. \cite{PRB-77-214403}, we can see a small jump near 100 K in the value of the spin relaxation time. It is, therefore, tempting to suggest that the discrepancy seen by Sutter \textit{et al.} \cite{PRB-75-140402} and Kitagawa \textit{et al.} \cite{PRB-77-214403} may be related to the structural deformation we observed at 110 K.

\subsection{Specific heat}

\begin{figure}
\begin{center}
\leavevmode
\includegraphics[width=0.7\textwidth]{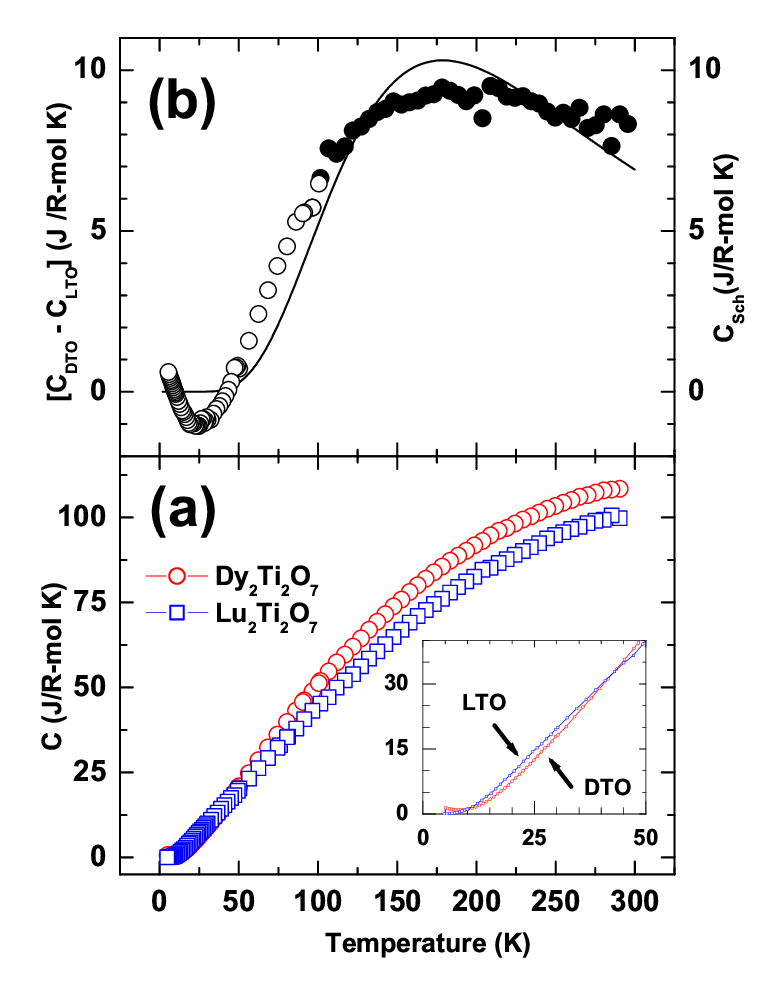}
\caption{(Color online) (a) Temperature dependence of specific heat (C) for the pyrochlores $Dy_2Ti_2O_7$ (DTO) and $Lu_2Ti_2O_7$ (LTO) in the temperature range 5 K to 300 K. Inset: blown-up view of the low temperature part of the main panel; (b) closed and open circles are data points corresponding to the difference $C_{DTO} - C_{LTO}$. The closed circles represent approximate Schottky contribution (C$_{Sch}$) in the specific heat of DTO above 100 K. The solid line is the calculated Schottky anomaly due to the first five higher lying crystal field doublets at 385, 503, 547, 679 and 956 K taken from ref.  \cite{PRB-77-214403}.} 
\label{Fig:7}
\end{center}
\end{figure}

The specific heat of $Dy_2Ti_2O_7$ (C$_{DTO}$) and $Lu_2Ti_2O_7$ (C$_{LTO}$) single crystals in the temperature range T = 5 to 300 K is shown in Fig. \ref{Fig:7}(a). Specific heat of both compounds shows a smooth monotonic decrease upon cooling below room temperature. For $Dy_2Ti_2O_7$, a tendency towards increase in the specific heat values when cooled below 10 K is due to short-range magnetic correlations between the Dy$^{3+}$ spins \cite{Nature-399-333}. In particular, no anomalous feature is seen in C$_{DTO}$ around T = 100 K, the temperature where the ``new'' mode first appears in Raman spectrum. While the specific heat of non-magnetic $Lu_2Ti_2O_7$ arises purely due to thermal lattice vibrations, the specific heat of $Dy_2Ti_2O_7$ has an additional Schottky contribution arising from thermal (de)population of higher lying CF split levels of Dy$^{3+}$ ions. Therefore, C$_{DTO}$ above 10 K can be expressed as an algebraic sum of contributions due to Schottky term (C$_{Sch}$) and thermal lattice vibration term (C$_{latt}$). Since $Dy_2Ti_2O_7$ is structurally analogous to $Lu_2Ti_2O_7$, albeit with a slightly smaller molecular weight, C$_{latt}$ term in the specific heat of $Dy_2Ti_2O_7$ can be approximated by the measured specific heat of non-magnetic $Lu_2Ti_2O_7$. However, as shown in the inset of Fig. \ref{Fig:7}(a), C$_{LTO}$ in the range T = 10 K to 40 K exceeds C$_{DTO}$ by values up to 1 J/R-mol K ($\sim$ 10 \% near T = 25 K). We note that this difference remains large negative even after rescaling C$_{LTO}$ by a factor  $\mu = (M_{DTO}/M_{LTO})^n$ (for both n = 1/2 and 3/2), where M$_{DTO}$ and M$_{LTO}$ are molecular weights of $Dy_2Ti_2O_7$ and $Lu_2Ti_2O_7$, respectively. These observations suggest that the lattice specific heat of $Dy_2Ti_2O_7$ is smaller than the specific heat of non-magnetic $Lu_2Ti_2O_7$, which is possibly due to the thermal lattice expansion behavior of these compounds shown in Fig. \ref{Fig:6}. While the lattice of $Lu_2Ti_2O_7$ expands continuously upon heating above 10 K, $Dy_2Ti_2O_7$ exhibits a relatively small lattice expansion at first until it expands steeply between 80 and 100 K and thereafter in a manner analogous to that of $Lu_2Ti_2O_7$. What transpire from the above discussion is that $Lu_2Ti_2O_7$ in not a useful lattice template for $Dy_2Ti_2O_7$ at low temperatures. However, above T = 100 K the thermal expansion behavior and the degree of anharmonicity (vide infra) of the two compounds are in quantitative agreement, therefore, the specific heat of $Lu_2Ti_2O_7$ is used to approximate the lattice specific heat of $Dy_2Ti_2O_7$ in the temperature range 100 to 300 K. A rough estimate of the Schottky contribution in the specific heat of $Dy_2Ti_2O_7$ above T = 100 K is obtained using: $C_{Sch} (= C_{DTO} - C_{LTO})$ . C$_{Sch}$ obtained using this procedure is shown in Fig. \ref{Fig:7}(b) by closed circles. It exhibits a broad maximum centered at T = 175 K with a value of $\sim$ 9 J/Dy-mol K. For comparison, we calculated the Schottky specific heat using the crystal field levels scheme given in ref. \cite{PRB-77-214403}, estimated by rescaling the CF parameters of $Ho_2Ti_2O_7$ obtained by Rosenkraz \textit{et al.} \cite{JAP-87-5914} using inelastic neutron scattering data. The calculated Schottky anomaly exhibits a peak at roughly the same temperature as the experimentally derived C$_{Sch}$, with a value of about 10 J/R-mol K at the peak temperature of 175 K. On the other hand, the calculated Schottky anomaly by Jana \textit{et al.} \cite{JMMM-248-7} (shown in Fig. 2 of ref. \cite{JMMM-248-7}) exhibits a huge peak of 9.5 J/Dy-mol K at a much lower temperature of T = 65 K. Moreover, at still lower temperatures, where C$_{DTO}$ is less than C$_{LTO}$ and hence C$_{Sch}$ is expected to be very small, the calculated C$_{Sch}$ due to Jana \textit{et al.} ranges from about 1 J/R-mol K near T = 20 K to 6.5 J/R-mol K near T = 40 K. Our specific heat data clearly favors the crystal field levels scheme of Rosenkranz \textit{et al.}, while rule out the validity of the crystal field scheme due to Jana \textit{et al.} \cite{JMMM-248-7}. From the position of maximum at 175 K in the experimentally derived C$_{Sch}$, it is easy to appreciate that the first excited crystal field split level of Dy$^{3+}$ in $Dy_2Ti_2O_7$ lies well above room temperature as proposed by Rosenkranz \textit{et al.} \cite{JAP-87-5914}. In Fig. \ref{Fig:7}(b) the open circles represents the difference $C_{DTO} - C_{LTO}$ below T = 100 K which disagree with the calculated C$_{Sch}$ curve, because at low temperatures, use of  $Lu_2Ti_2O_7$ as a lattice template slightly overestimates the C$_{latt}$ of $Dy_2Ti_2O_7$ leading to even negative values of the difference $C_{DTO} - C_{LTO}$ below T = 40 K as discussed above.  

In short, the specific heat investigations of $Dy_2Ti_2O_7$ and $Lu_2Ti_2O_7$ clearly indicate that the position of first excited crystal field split doublet is close to 350 K (240 cm$^{-1}$). These results favor the crystal field scheme for $Dy^{3+}$ in $Dy_2Ti_2O_7$ derived from the work of Rosenkranz \textit{et al.} on $Ho_2Ti_2O_7$ \cite{JAP-87-5914}, however, due to problems associated with correct determination of C$_{Sch}$ at low temperatures, the validity of crystal field splitting due to Rosenkranz \textit{et al.} could not be verified completely.

\subsection{Anomalous red-shift of phonons}

\begin{figure}
\begin{center}
\leavevmode
\includegraphics[width=0.9\textwidth]{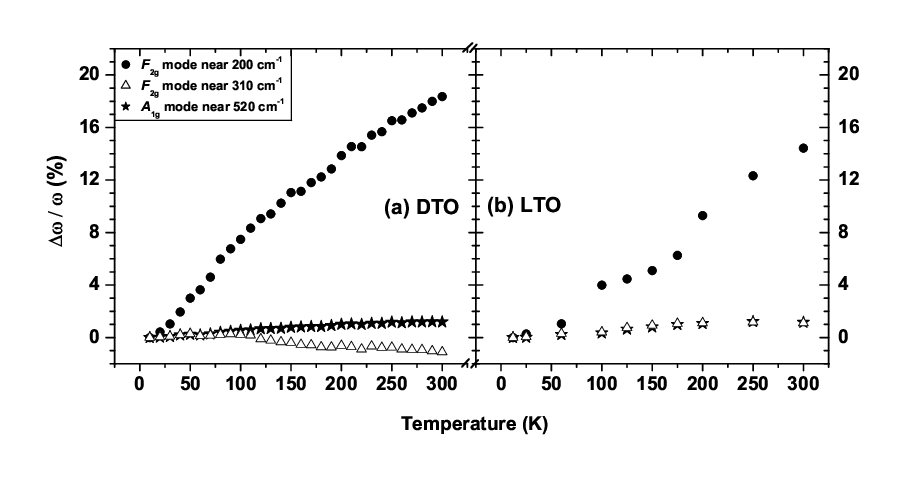}
\caption{Percentage change of frequencies ($\frac{\Delta\omega}{\omega}$\%) of (a) $Dy_2Ti_2O_7$ and (b) $Lu_2Ti_2O_7$ phonons with temperature which thus indicates a similar amount phonon softening in both the pyrochlores.} \label{Fig:8}
\end{center}
\end{figure}

For spin-ice $Dy_2Ti_2O_7$, as shown in Fig. \ref{Fig:3}, a large anomalous softening (\textit{red-shift}) of the modes M1, M3, M4, M9 and M11 is seen with decreasing temperature. The new mode (M5) that appears below 110 K also shows a considerable large amount of \textit{red-shift} with decreasing temperature. Interestingly, the phonon modes of the non-magnetic $Lu_2Ti_2O_7$ also show large anomalous softening, as temperature decreases. As shown in Fig. \ref{Fig:4}, the phonons N1, N2, N3, N4, N6, N7 and N9 show \textit{red-shifts}, similar to the $Dy_2Ti_2O_7$ phonons. Temperature dependence of the strong $F_{2g}$ modes (M3, N3, M6, and N6) and the $A_{1g}$ mode (M9 and N9) of the two pyrochlores are compared in Fig. \ref{Fig:8} which emphasizes that magnitude of phonon softening with decreasing temperature in both pyrochlores is comparable.

Temperature variation of a phonon mode ($i$) of frequency $\omega_i$ can be expressed as,
\begin{eqnarray}
\omega_i(T)=&& \omega_i(0)+{(\Delta \omega_i)}_{qh}(T)+{(\Delta \omega_i)}_{anh}(T) \nonumber \\
&& +{(\Delta \omega_i)}_{el-ph}(T)+{(\Delta \omega_i)}_{sp-ph}(T)
\end{eqnarray}
The first term corresponds to the phonon frequency at T = 0 K. The second term is a contribution to the phonon frequency due to a change in lattice constant with temperature and is known as quasi-harmonic contribution to the frequency change. The third term corresponds to the intrinsic anharmonic contribution to phonon frequency that comes from the real part of the self-energy of the phonon decaying into two phonons (cubic anharmonicity) or three phonons (quartic anharmonicity). The fourth term ${(\Delta \omega_i)}_{el-ph}$ is due to coupling of phonons with charge carriers within the system which is absent in insulating pyrochlore titanates. The last term, ${(\Delta \omega_i)}_{sp-ph}$, is the change in phonon frequency due to spin-phonon coupling arising from modulation of spin exchange integral by the lattice vibration.

\begin{figure}
\begin{center}
\leavevmode
\includegraphics[width=0.7\textwidth]{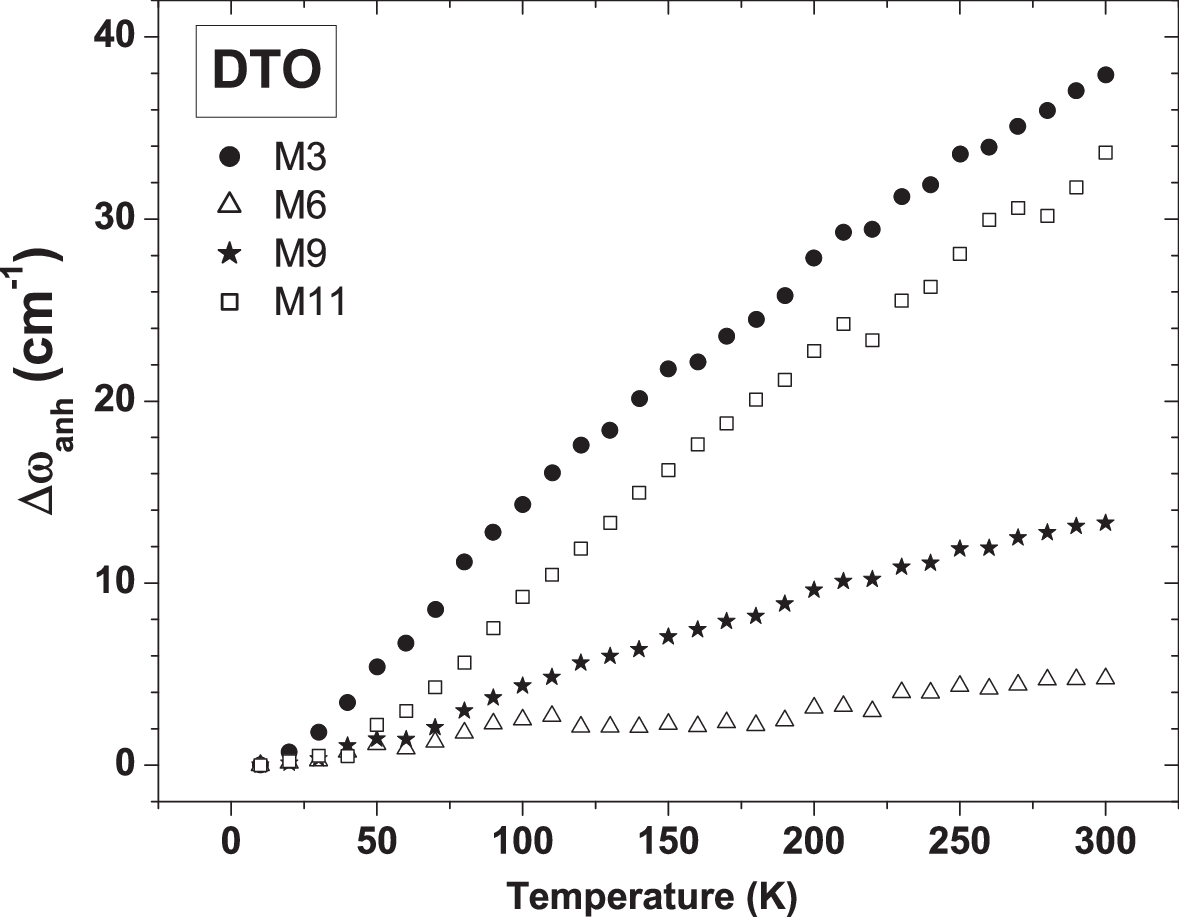}
\caption{The intrinsic anharmonic contribution ($\Delta\omega_{anh}$) to the change in frequency of $Dy_2Ti_2O_7$ phonons estimated from experimental data.} \label{Fig:9}
\end{center}
\end{figure}

The lattice contribution ${(\Delta \omega_i)}_{qh}(T)$ to the phonon energy accounts for an expansion / contraction of the lattice leading to a change in the harmonic force constant without changing the phonon population. This change in frequency of a mode $i$ can thus be approximately related to the change in volume using the Gr\"{u}neisen parameter $\gamma_i=-\frac{\partial \ln \omega_i}{\partial \ln V}=\frac{B_0}{\omega_i}\frac{\partial \omega_i}{\partial P}$, where $B_0$ is the bulk modulus. Taking T=12 K as the reference temperature (the lowest temperature in our measurements), $\frac{\omega_i(T)-\omega_i(12K)}{\omega_i(12K)}=-\gamma_i \frac{V_i(T)-V_i(12K)}{V_{12K}}$ \cite{ JRS-39-537}. In order to account for this quasi-harmonic contribution, we have taken the measured values of the Gr\"{u}neisen parameters from our high pressure Raman and x-ray experiments on $Dy_2Ti_2O_7$ at room temperature \cite{DTOhp} and the values are $\gamma$(M3)=2.2, $\gamma$(M6)=1.7, $\gamma$(M9)=0.9 and $\gamma$(M11)=2.2. The volume of the unit cell at different temperatures has been taken from the temperature dependence of the lattice parameters shown in Fig. \ref{Fig:6}. The quasi-harmonic contribution is subtracted from the experimental frequencies to obtain the change in frequency ($\Delta\omega$) solely due to intrinsic anharmonic effects (${(\Delta \omega_i)}_{anh}(T)$) and to the spin-phonon coupling (${(\Delta \omega_i)}_{sp-ph}(T)$). This is shown for the modes M3, M6, M9 and M11 in Fig. \ref{Fig:9}. 

The effect of spin-phonon coupling on the phonon frequency may be accounted for using \cite{JPCM-17-5225, PRB-73-052301, PRB-60-11879} ${(\Delta \omega_i)}_{sp-ph}(T)=\lambda<\bf{S}_i\bf{S}_j>$, where $<\bf{S}_i\bf{S}_j>$ is the spin correlation function and $\lambda$ is the spin-phonon coupling coefficient. In the temperature-dependent infrared studies of $Dy_2Ti_2O_7$ in ref \cite{JPCM-17-5225} the phonon anomalies have been attributed to the spin-phonon coupling. Our quantitative measurements of the phonon frequency for various modes in the spin-ice $Dy_2Ti_2O_7$ (Fig. \ref{Fig:8}(a)) and the non-magnetic $Lu_2Ti_2O_7$ (Fig. \ref{Fig:8}(b)) clearly show that the magnitude of phonon softening is comparable in both pyrochlores. Since $Lu_2Ti_2O_7$ is a non-magnetic pyrochlore, there is no contribution from ${(\Delta \omega_i)}_{sp-ph}(T)$ which thus rules out this mechanism for explaining the phonon anomalies in $Dy_2Ti_2O_7$. Therefore, the change in frequency ($\Delta\omega$) (as shown in Fig. \ref{Fig:9}) is solely due to intrinsic anharmonic effects (${(\Delta \omega_i)}_{anh}(T)$).

The anharmonic effects become considerable when the system temperature increases and a phonon decay into two or three phonons due to cubic or quartic anharmonic interactions. We recall that in the language of quantum many-body theory of phonons in insulating solids without isotopic mass disorder are described as quasi-particles with a complex self-energy: $\Sigma=\Delta + i\Delta^{\prime}$ originating from anharmonic phonon-phonon interactions. The complex self-energy is specific to each phonon and depends on frequency and temperature. The observed frequency $\omega_i(T)$ and damping parameter $\gamma_i$ exhibit a temperature dependence \cite{PRB-7-2779, PRB-28-1928, SSC-117-201}: 

\begin{eqnarray}
{\omega_i}^2={\omega_i}^2(0)+ 2 {\omega_i}(0) \Delta(\omega_i,T) \nonumber \\
i.e., \omega_i \approx \omega_i(0)\left[1+\frac{\Delta}{\omega_i(0)}\right] \\
and \hspace{5mm} \omega_i \gamma_i = 2 {\omega_i}(0) \Delta^{\prime}(\omega_i,T)
\end{eqnarray}
where $\omega_i(0)$ is the temperature-independent $i$th mode frequency in the harmonic approximation. Since the highest temperature in our experiment is 300 K, much smaller than the Debye temperature of the titanate pyrochlores ($\sim$ 500 K), it is sufficient to consider only the contribution of cubic anharmonicity. Considering only the simplest decay channels, as was done by Klemens \cite{PR-148-845}, phonon of frequency $\omega(\vec{q}\sim \vec{0})$ decays into two phonons $\omega_1(\vec{q}_1)$ and $\omega_2(\vec{q}_2)$ keeping the phonon energy and momentum conserved, i.e. $\omega_i(0) = \omega_1 + \omega_2$ and $\vec{0}=\vec{q}_1 + \vec{q}_2$, then,

\begin{equation}
\Delta_i(\omega_i,T)={\Delta_i}^A[1+2\bar{n}(\omega_i(0)/2)]
\end{equation}
where $\bar{n}$ is the Bose-Einstein mean occupation number of the $i$th state. The self-energy parameter ${\Delta_i}^A$ can be positive or negative. Most often, the value of ${\Delta_i}^A$ is negative, i.e., the observed mode frequency decreases with increasing temperature, which is the ``regular'' behavior. However, in case of phonons termed ``anomalous'' the sign of  ${\Delta_i}^A$ has to be positive, which is most likely the case for the titanate pyrochlores. Beside the titanate pyrochlores, other examples of anomalous phonons include: $B_{1g}$ Raman-active phonon in $SnO_{2}$ \cite{PRB-7-2779}, $B_{1g}$ Raman-active phonons in $Pr_{1.85}Ce_{0.15}CuO_4$ (see Fig. 7 of ref. \cite{PRB-43-2857}), and infrared-active $TO_3$ and $LO_4$ phonons in spinels $MgAl_2O_4$ (see Fig. 3 of ref. \cite{PRB-73-064305}) and, the one which is most relevant to the present study, $F_{2g}(1)$, $F_{2g}(2)$ and $E_g$ Raman-active modes of ${\beta}$-pyrochlore $KOs_2O_6$ \cite{PRB-77-064303}. Here, the phonon anomaly has been found not only for the $F_{2g}(1)$ mode involving potassium, but also for two oxygen modes $F_{2g}(2)$ and $E_g$, arising from highly anharmonic vibrations of oxygen ions. We have discussed the relelvance of this to the present case in the next paragraph. It has been argued \cite{SSC-117-201} that the sign of ${\Delta_i}^A$ is negative or positive depending on whether the restricted two-phonon density of states $\tilde{D}(\omega)$ (integrated over all decay channels $\omega_i \rightarrow \omega_1+\omega_2$ and $\vec{0} = \vec{q}_1+\vec{q}_2$) has a maximum at a frequency higher or lower than the phonon frequency. It will be very interesting to obtain $\tilde{D}(\omega)$ for $Dy_2Ti_2O_7$ from \textit{ab initio} lattice dynamical calculations to explain the anomalous temperature dependance quantitatively. 

We wish to speculate why the phonon-phonon anharmonic interactions are strong in the titanate pyrochlores. The most anomalous phonon is the $F_{2g}$ mode near 200 cm$^{-1}$, which is related to O$^{\prime}$ sublattice vibration. We recall that the 8b site in the pyrochlore structure are occupied by O$^{\prime}$ ions. However, the 8a sites are vaccant. The presence of unoccupied 8a sites can result in large vibrational amplitude of O$^{\prime}$ ions and hence make $F_{2g}$ mode at 200 cm$^{-1}$ most anomalous. In order to test the validity of our assumtion, we did a very rough comparision of the displacement parameter $<u^2>$ of oxygen ions in ${\beta}$-pyrochlore $KOs_2O_6$ taken from the single-crystal x-ray analysis of ref. \cite{JSSC-179-336} with that in the pyrochlore $Tb_2Ti_2O_7$ for which $<u^2>$, obtained from neutron diffraction data, is available between 4.5 K and 600 K \cite{PRB-69-024416}. We find that the ratio of square-root of vibrational amplitude of oxygen to the bond length $d_{A-O}$ (A = K or Tb), for the two compounds, have nearly the same magnitude ($\sim 3\%$). Further, O ion at 48f sites are also slightly displaced from their ideal position $\frac{3}{8}a$ to $(\frac{3}{8}-x)a$, where a is the lattice parameter \cite{PSSC-15-55}. The will make the vibrational amplitude of O ion at 48f sites larger and hence the $A_{1g}$ mode at 519 cm$^{-1}$ will also be more anharmonic than usual but less than the 200 cm$^{-1}$ $F_{2g}$ mode involving O$^{\prime}$ ion.

\section{Summary and Conclusion}

In summary, the Raman light scattering properties of the spin ice $Dy_2Ti_2O_7$ and its non-magnetic analogue $Lu_2Ti_2O_7$ have been studied in detail between T = 12 to 300 K. Our observations are as follows: (1) In $Dy_2Ti_2O_7$ crystal a ``new'' mode not predicted by factor group analysis of Raman-active modes for the pyrochlore lattice emerges upon cooling the sample below T = 110 K. Raman data from O$^{18}$-isotopic $Dy_2Ti_2O_7$ strongly indicate the phononic nature of this mode. On the other hand, in $Lu_2Ti_2O_7$, no ``new'' mode appears in the whole temperature range studied; (2) the slope ($\frac{d\omega}{dT}$) for the modes M1 and M7, shows a distinct change of sign at T = 110 K, (3) a large anomalous softening of some of the phonons with decreasing temperature is seen in both $Dy_2Ti_2O_7$ and $Lu_2Ti_2O_7$; (4) at the same temperature, the cubic lattice parameter of $Dy_2Ti_2O_7$ changes abruptly by about 0.1 \%, while the lattice parameter of non-magnetic $Lu_2Ti_2O_7$ varies smoothly over the entire temperature range of our investigations.  

The appearance of a ``new'' (phonon) mode in the Raman specta is indicative of symmetry lowering in $Dy_2Ti_2O_7$ below T = 110 K. However, the powder x-ray diffraction peaks of $Dy_2Ti_2O_7$ do not show any spliting or emergence of new Bragg peak at least down to about T = 10 K. Thus, within the experimental limits of our powder x-ray diffraction experiments, the symmetry lowering in $Dy_2Ti_2O_7$, if any, could not be detected. It should be recalled that all \textit{six} Raman active phonon modes in the pyrochlore lattice are due to vibrations of the oxygen ions alone, rare-earth and titanium ions being located at the centre of inversion. Since, oxygen ion has a very low x-ray scattering cross-section, a ``subtle'' change in the anionic sublattice may go unnoticed in x-ray powder diffraction experiments. We, therefore, propose that a ``subtle'' structural deformation occurs in the anionic sublattice of $Dy_2Ti_2O_7$ near $T_c = 110$ K, giving rise to a ``new'' phonon mode in the Raman spectra below this temperature. A similar situation has recently been noted by Knee \textit{et al.} \cite{PRB-71-214518} for the pyrochlore superconductor $Cd_2Re_2O_7$ which exhibits a non-trivial structural deformation involving a small deformation in the $ReO_6$ octahedral lattice, at 200 K, giving rise to some additional low frequency Raman modes. 

Finally, our study of the non-magnetic analogue $Lu_2Ti_2O_7$ which shows an anomalous softening of phonons of comparable magnitude as that in $Dy_2Ti_2O_7$ suggests that spin-phonon coupling is not responsible for the anomalous temperature dependence of phonons in these and other members of the rare-earth titante pyrochlore series. We rather propose that strong phonon-phonon anharmonic interaction can lead to these anomalous phonons. 

To conclude, Raman spectroscopy data, in conjunction with x-ray diffraction studies of the spin-ice $Dy_2Ti_2O_7$ and of its non-magnetic analogue $Lu_2Ti_2O_7$ are reported over a largely unexplored temperature window, between 10 to 300 K. Our data reveal a ``subtle'' structural deformation of the $Dy_2Ti_2O_7$ lattice at 110 K. We show that anomalous phonon softening is a generic feature of rare-earth titanate pyrochlores which is not limited to the magnetic members of this family, thereby ruling out the spin-phonon coupling in these pyrochlores which was earlier proposed \cite{JPCM-17-5225} to be responsible for the anomalous phonon behavior. We propose that this behavior is due to unusually strong phonon anharmonicity in these pyrochlores. We believe that our experimental results will motivate further studies on the spin ice $Dy_2Ti_2O_7$ in order to understand the nature of the phase transition near 110 K and its possible consequences on the spin ice models.

\begin{acknowledgments}
We thank the Indo-French Centre for Promotion of Advanced Research (IFCPAR), Centre Franco-Indien pour la Promotion de la Rechereche Avanc\'{e}e (CEFIPRA) for financial support under Project No. 3108-1.  AKS thanks also the Department of Science and Technology (DST), India for financial support.
\end{acknowledgments}

\newpage

\begin{table}
\caption{\label{tab:table2}A list of the experimental frequencies in cm$^{-1}$ at 12 K.}
\begin{ruledtabular}
\begin{tabular}{lc}
 Normal modes                      &Experimental \\
\hline 
 M1 \footnotemark[1] (Phonon)      &93           \\
 M2 \footnotemark[1] (Phonon)      &126          \\
 M3 (Phonon, $F_{2g}$)             &174          \\
 M4 \footnotemark[1] (Phonon)      &194          \\
 M5 (New phonon mode below 110 K)  &287          \\
 M6 (Phonon, $F_{2g}$)             &312          \\
 M7 (Phonon, $E_{g}$)              &330          \\
 M8 (Phonon, $F_{2g}$)             &453          \\
 M9 (Phonon, $A_{1g}$)             &515          \\
 M10 (Phonon, $F_{2g}$)            &563          \\
 M11 \footnotemark[1] (Phonon)     &680          \\
 M12 \footnotemark[1] (Phonon)     &712          \\
 M13 \footnotemark[1] (Phonon)     &791          \\
 M14 \footnotemark[1] (Phonon)     &867          \\
\end{tabular}
\end{ruledtabular}
\footnotetext[1]{Origin of the mode has been discussed in the text.}
\end{table}

\newpage


\end{document}